\documentclass[aps,superscriptaddress,showpacs,nofootinbib,eqsecnum,pre,notitlepage]{revtex4-1} 

\usepackage{amssymb}   
\usepackage{amsmath}   
\usepackage{natbib}
\usepackage{epsfig}   
\usepackage{graphicx}   
\usepackage{dcolumn}
\usepackage{bm}  
\usepackage[english]{babel}
\usepackage[latin1]{inputenc}
\usepackage{eucal} 
\usepackage{verbatim}
\usepackage{latexsym}
\usepackage[update]{epstopdf}
\begin{document}

\title{Colored noise and memory effects on formal spiking neuron models}

\author{L. A. da Silva} \email{leandro.silva@ufabc.edu.br}
\affiliation{Centro de Matemática, Computação e Cognição, UFABC, Santo André-SP, Brazil}
  
\author{R. D. Vilela} \email{rafael.vilela@ufabc.edu.br}
\affiliation{Centro de Matemática, Computação e Cognição, UFABC, Santo André-SP, Brazil}

\begin{abstract}

Simplified neuronal models capture the essence of the electrical activity of a generic neuron, besides being more interesting from the computational point of view when compared to higher dimensional models such as the Hodgkin-Huxley one. In this work, we propose a generalized resonate-and-fire model described by a generalized Langevin equation that takes into account memory effects and colored noise. We perform a comprehensive numerical analysis to study the dynamics and the point process statistics of the proposed model, highlighting interesting new features like: i) non-monotonic behavior (emergence of peak structures, enhanced by the choice of colored noise characteristic time-scale) of the coefficient of variation (CV) as a function of memory characteristic time-scale, ii) colored noise-induced shift in the CV,  and iii) emergence and suppression of multimodality in the interspike interval (ISI) distribution due to memory-induced subthreshold oscillations. Moreover, in the noise-induced spike regime, we study how memory and colored noise affects the coherence resonance (CR) phenomenon. We found that for sufficiently long memory, CR is not only suppressed, but also the minimum of the CV $\times$ noise intensity curve that characterizes the presence of CR may be replaced by a maximum. The aforementioned features allow to interpret the interplay between memory and colored noise as an effective control mechanism to neuronal variability. Since both variability and non-trivial temporal patterns in the ISI distribution are ubiquitous in biological cells, we hope the present model can be useful in modeling real aspects of neurons.

\end{abstract}


\maketitle

\section{Introduction}
\label{sec1}

The current consensus is that neurons are the fundamental units where information is processed  in the nervous system. The celebrated Hodgkin-Huxley (HH) model mathematically translates into a four dimensional ordinary differential equation the detailed biophysical features of a neuron and nicely explains the generation of an action potential. Besides its relatively high dimensionality,  another downside of the HH model rests on the fact that there are   approximately twenty parameters to be determined. Consequently, a clear and intuitive understanding of the fundamentals of neuronal dynamics becomes difficult and one is often restricted to computer simulations. Also, this high dimensionality implies high computational costs when one aims at simulating a large population of neurons. 

If an accurate low-dimensional description of neuron activity could be found, it would certainly have advantages over high dimensional approaches, both in terms of simulation and understanding of the fundamental principles. In addition, the understanding of neuronal mechanisms resorting to simple models could be useful in the engineering of artificial neural devices designed to reproduce a given real biological feature. In that sense, several alternative approaches have been created along the years, like the spike-response class of models. Two notable examples in this class are the integrate-and-fire  \cite{Burkitt2006} and the resonate-and-fire models  \cite{Izhikevich2001}. These kinds of models are typically characterized  by a differential equation whose solution describes the time evolution of the subthreshold membrane potential and an \textit{ad hoc} rule according to which when the potential reaches the threshold (i) a spike is produced and (ii) the potential is reset to a certain value. Additionally, a refractory period, that is natural in the HH model, can be postulated in the such models.

The precursor of integrate-and-fire model was introduced by the French physiologist Louis Lapicque in 1907, long before the introduction of the HH model \cite{lapicque1907recherches}. In 1965, Richard Stein introduced the so-called \textit{leaky integrate-and-fire} (LIF) model \cite{stein1965theoretical}, which became very popular from the 1990s, when it started to be used (along side with its generalizations) in neural networks studies \cite{Brunel2007,sacerdote2013stochastic}. Essentially, the integrate-and-fire model is constructed based on analogies with electrical circuits. We can picture the standard LIF model as a parallel resistive-capacitive (RC) circuit subjected to a deterministic external input. In addition to this deterministic input, there are several sources of fluctuation that influence the dynamics of the neuron potential. We can name at least three of these sources that stand out. The opening and closing of the ion channels behave stochastically \cite{hille2001ion}, the release of neurotransmitters by chemical synapses is a Poissonian process, and the synaptic input from other neurons (of the order of $10^4$ synaptic junctions per neuron \cite{rolls2010noisy}) can also be well modeled as a stochastic process \cite{Lindner2009}.
Stochastic versions of neuron models have been strongly motivated by \textit{in vivo} studies, since in this case the random time intervals of input synaptic impulses have to be taken into account. A prominent example in this class is the diffusive (or stochastic) leaky integrate-and-fire (SLIF). The main characteristic of the SLIF model is the explicit description of the randomness of synaptic inputs through the presence of a noise term in the equation of motion of the membrane potential $v$. 

Far from being just a mathematical device to dictate how the environment manifests itself in the effective dynamics of the system, the presence of noise in neural dynamics has proven fundamental in neurophysiological processes \cite{rolls2010noisy}. Experimental evidence supports the idea that noise is intrinsically necessary to mechanisms responsible for signal detection, decision making, perception and short term memory  \cite{rolls2010noisy,Deco2009a,Deco2010,Deco2008}. Therefore, detailed studies connecting stochastic processes and neuronal dynamics at all levels (i.e. single neurons, networks and continuum field approaches) are necessary for a better understanding of the complexity of the nervous system.

The subthreshold dynamics of the SLIF model is phenomenologically described by the following stochastic differential equation, already written in the dimensionless form~\cite{Burkitt2006}:     
\begin{equation} \label{SLIF}
\tau \frac{dv(t)}{dt} + v(t) =  \mu + \sqrt{2\sigma^2} \eta(t)\;, 
\end{equation}
where $v(t)$ is the membrane potential at time $t$,  $\tau$ corresponds to the membrane time scale and $\mu$ is a constant average input if synaptic inputs are considered homogeneous. 
The resistor plays the role of the Ohmic leakage impedance of the neural membrane and the capacitor models the lipid bilayer that forms the neuronal membrane. In Eq.~(\ref{SLIF}), $\eta(t)$ is to be interpreted as a Gaussian white noise term (i.e. $\langle \eta(t) \rangle = 0$ and $\langle \eta(t) \eta(t') \rangle = \delta(t-t')$) and $\sigma$ is the noise intensity.  

Eq.~(\ref{SLIF}) corresponds to the simplest possible formulation of the SLIF model and may lose sight of some relevant biological features. In view of this, several generalizations have been proposed in the literature, many of them supported by experimental evidence. For example, Stevens and Zador suggest that the variability of output spike times of cortical neurons in the awake brain could be accounted for the existence of correlations in the arrival times of input at different synapses \cite{stevens1998input}. In Ref.~(\cite{Sejnowski2002}), Salinas and Sejnowski highlight the necessity to take into account the impact of temporal correlations. Their motivation comes from experimental studies showing that cortical neurons can exhibit an input correlation time of the same order of magnitude as the mean interspike interval of the correspondent response. A practical way to implement a finite correlation time is to promote the term $\eta(t)$ in Eq.~(\ref{SLIF}) to a colored noise term.

Many other studies have been developed in order to achieve a more realistic (yet effective and simplified) approach to neuronal dynamics \cite{Linkenkaer-hansen2004,Svirskis2000,Masoller2008,muller2015statistics}, and we would like to go a step forward. In a broader perspective, we can draw an analogy between neuronal dynamics and the nonequilibrium statistical theory of open systems. We can interpret a given neuron (or even a given neuronal tissue) in whose dynamics we are interested as the ``system''. Since this system interacts with its surrounding medium, we have to account for the environment feedback into the system in order to obtain an accurate description of the system dynamics.  In the early times of the development of the theory of open systems, the usual approach used to describe the effective dynamics of these systems was to model the problem resorting to a phenomenological Langevin equation. The original Langevin equation is characterized by Gaussian white noise and time-local dissipative terms. The main limitation of this approach is that the usual Brownian motion implicitly assumes that the environment interacts instantaneously with the system. This assumption is valid only in very restricted and particular limit cases. If a more realistic description of a given open system is required, it is necessary to take into account that information exchange (e.g. linear momentum exchange between massive particles, ionic current exchange between neurons and so on) takes place during finite time intervals, which must lead to finite memory effects and colored noise. In view of this, some phenomenological \cite{kubo1966fluctuation} and microphysical \cite{Caldeira,calzetta2003stochastic,hasegawa2011classical,gleiser1994microphysical} approaches  have been developed in order to elucidate how a more realistic effective equation of motion emerges when the environment degrees of freedom are coarse-grained. From this more fundamental perspective, a generalized theory of Brownian motion arose, which can be described by a generalized Langevin equation (GLE). In such an equation, nonlocal dissipative terms and correlated noise reflect finite-time effects.

In this paper, we argue that it would be a step toward a more realistic modeling of effective neuronal dynamics to rewrite the equation that describes the membrane potential using the generalized Brownian motion framework. We thus propose to enrich intrinsically phenomenological neuron models to reflect a crucial aspect that characterizes first-principles models: the \textit{concomitant} presence of memory effects and colored noise. This mathematical framework ensures causality and is particularly a requisite when the processes that take place in the system and environment have comparable characteristic time scales. We thus introduce and characterize a first-passage time model whose subthrehold dynamics obeys the Langevin  equation of generalized Brownian motion.

From the biophysical point of view, it is empirically known that the distribution of the interspike intervals (ISI) of real neurons (e.g. pallidal and ganglion cells) can exhibit non-trivial patterns, like a bimodal or multimodal structure depending on the input \cite{dorval2008deep,sirovich2014cooperative,levine1977model,worgotter1998influence}. Neither the standard nor the stochastic LIF model is able to produce such a multimodal distribution. Even though the original resonate-and-fire model is capable of providing multimodality under some conditions \cite{Verechtchaguina2006,chacron2007threshold,coombes2011neuronal}, it does not have an intrinsic dynamical mechanism to control the emergence and the shape of multimodality. We show that the memory term introduced by our model plays the role of a variability control mechanism. Adjusting the memory parameter, which reflects a property of the input, one enables smooth transitions between unimodal and multimodal regimes.

The paper is organized as follows: in Sec.~\ref{sec3} we introduce our model based on a first passage time (FPT) problem for the generalized Brownian motion, and review some analytical results on the subthreshold dynamics to guide the understanding of the numerical analysis, presented in Sec~\ref{numerical}. The results and discussions are the subject of Sec~\ref{results}. In Sec.~\ref{outline}, the outlook of this paper is given.

\section{A generalized resonate-and-fire model}
\label{sec3}

The SLIF model is an useful tool to mimic the potential dynamics of the so-called nonresonant cells. The response of this kind of cell to an external excitatory stimulus is characterized by a monotonic relaxation toward its resting state. This behavior becomes even clearer if we draw an analogy between Eq.~(\ref{SLIF}) and the equation of motion of an overdamped harmonic oscillator. In recent decades, a substantial amount of experimental data has shown that neurons can also exhibit damped subthreshold oscillations at a given frequency \cite{erchova2004dynamics,wu2001membrane,klink1993ionic,liu2000spinal,fransen2004ionic}. From a dynamical viewpoint, this occurs when the neurons operate close to an Andronov-Hopf bifurcation \cite{izhikevich2007dynamical}. The specific case in which there is no bistability and the rest state is the global attractor corresponds to a supercritical Andronov-Hopf bifurcation. Several models that aim at accounting for the aforementioned subthreshold oscillations have been constructed \cite{Verechtchaguina2006,erchova2004dynamics,georgiev2003identifying,izhikevich2003simple}.  Our starting point is an archetypal second-order model~\cite{Verechtchaguina2006} for the subthreshold dynamics, written in dimensionless form as:
\begin{equation} \label{RIF}
\lambda \frac{d^2v(t)}{dt^2} + \gamma \frac{dv(t)}{dt} + \omega^2 v(t) = \mu  + \sqrt{2\sigma^2} \eta(t) \;,
\end{equation}
where $\eta(t)$ is again taken as white Gaussian noise. In terms of the analogy between neuronal dynamics and electrical circuits,  the second-order derivative in the left hand side of Eq.~(\ref{RIF}) emerges when an inductor is plugged in parallel with both $R$ and $C$ in the LIF circuit. From the physiological point of view, the inductor emulates the presence of long-term processes, like calcium currents \cite{nicoll1988current,mahanty1998calcium}. An important feature of the model given by Eq.~(\ref{RIF}) is that for $\gamma < 2\omega\sqrt{\lambda}$ (underdamped case) the typical behavior of a resonant neuron is recovered. Analogously, $\gamma > 2\omega\sqrt{\lambda}$ corresponds to the overdamped case. In particular, in the limit case when $\lambda=0$ (usually referred to as the overdamped limit) the SLIF neuron that mimics the behavior of a nonresonant cell is recovered. Therefore, Eq.~(\ref{RIF}) can be thought of as a generalization of the SLIF model. Here we are interested in the case $\lambda \neq 0$.

A more general approach corresponds to rewriting Eq.~(\ref{RIF}) using the framework of the generalized Brownian motion which, after rescaling the parameters in terms of $\lambda$, reads:
\begin{equation} \label{GLE}
\frac{d^2v(t)}{dt^2} + \gamma \int_{t_0}^{t} dt' K(t-t') \frac{dv(t')}{dt'} + g'(v(t)) =  \xi(t) \;, 
\end{equation}
where $t_0$ is the time instant when the reset rule was last performed,  $K(t-t')$ is the memory kernel whose  integral is normalized to the unit and whose characteristic time scale is denoted by $1/\Gamma$, and $\xi$ is a Gaussian colored noise with characteristic time scale $1/\Gamma_{\xi}$. The presence of colored noise is particularly relevant in situations where the characteristic time scales of fluctuations are comparable to the system time scale. The deterministic parameter $\mu$ in Eq.~(\ref{RIF}) was absorbed in $g'(v)$, which is a linear or nonlinear function of the membrane potential. The prime in $g'(v)$ means derivative w.r.t $v$. 
If the neuron worked strictly as a RLC circuit and if the only source of fluctuation were of thermal origin, then a fluctuation-dissipation relation (FDR) would be expected to hold exactly (Nyquist noise): $\langle\xi(t)\xi(t')\rangle=K(t-t')$. Here, as mentioned above, we are motivated by other sources of noise as well, and do not impose beforehand such a perfect matching between noise and memory time scales.

Equation~(\ref{GLE}) is the central equation of this paper. Once the memory kernel is defined (see Sec~\ref{numerical} for a convenient choice), it describes the subthreshold evolution of the membrane potential. The establishment of a fire-and-reset rule completes the model. Here we adopt the rule: if $v(t)=v_{\rm{th}}$ then (i) a spike is considered to have occurred at time $t$ and (ii) $v \rightarrow v_{\rm{r}}$ and $v' \rightarrow 0$. Generally speaking, the reset rule reflects the biophysical fact that subsequent to a spike is the return of all the neuron state variables (e.g. in the Hodgkin-Huxley model) to a small region in phase space. The reset rule corresponds to the simplification that replaces that small region by a single point. Since the equation for the subthreshold dynamics, Eq.~(\ref{GLE}), is of second-order for the variable $v$, it can be rewritten as a (two-dimensional) first-order equation for the variables $v$ and $dv/dt$. These are the state variables of our model and, accordingly, must be both reset. The specific choice of the value zero for the reset of $dv/dt$ does not have any special meaning and implies no loss of generality.

We can recover the standard first order SLIF model from Eq.~(\ref{GLE}) by applying two approximations. The first one is the Markovian approximation, whose validity tends to be assured when the time scale of the system $T \equiv v/\dot v$ is such that $T \gg 1/\Gamma$ (additionally, $T \gg 1/\Gamma_{\xi}$ if the white noise limit is also required), and for large times, when $t/T \gg 1$. The nonlocal term can be then rewritten as \cite{zwanzig1973nonlinear,hasegawa2011classical}
\begin{equation} \label{markapprox}
\gamma \int^{t}_{t_0} dt' K(t-t') \dot v(t') \approx  \gamma  \int^{t}_{t_0} dt'\; 2\delta(t-t') \dot v(t') = \gamma \dot v(t)\;. 
\end{equation}

At this point, we recover the second-order model Eq.~(\ref{RIF}) with $\lambda=1$,
which will be referred to as the Markovian approximation and compared to our proposed model Eq.~(\ref{GLE}) in Sec.~\ref{results}. Going a step forward and performing a second approximation by imposing $\lambda=0$, we recover the first order SLIF model. The term $g'(v)$  can be interpreted in mechanical terms as the derivative of an effective potential. For a linear choice of $g'(v)$, we can define $g(v) = \omega^2 v^2/2 - \mu v $, with minimum given by $v_{c} = \mu/\omega^2$. For any $v(t=0) < v_{\rm{th}} < v_{c}$, given a sufficiently long time interval, the probability of spike occurrence is 1, both in the deterministic and in the stochastic cases. This regime is usually known as  \textit{mean-driven} or \textit{tonic} firing regime. On the other hand, if  $v(t=0) <  v_{c} < v_{\rm{th}} $, the probability of occurrence of spikes in the overdamped regime is non-null only in the stochastic case. The same occurs in the underdamped regime if the maximum value of the time-dependent amplitude of oscillation around $v_c$ is smaller than $ v_{\rm{th}}$ when the noise is switched off. One refers to these cases as the \textit{fluctuation-} or \textit{noise-induced} firing regime.

If we consider a linear version of Eq.~(\ref{GLE}), standard Laplace transform techniques can be used to assess the subthreshold dynamics of the model \cite{adelman1976generalized}. This procedure will be useful as a check for the numerical code and also to assist the FPT analysis in the next sections, when the fire-and-reset rule is turned on.  Imposing a linear $g'(v)$ function defined as $g'(v)=\omega^2 v - \mu$ in Eq.~(\ref{GLE}), we obtain:
\begin{equation} \label{v_laplace}
\bar{v}(s) = \frac{\dot{v}(0)+\left[ s + \gamma  \bar{K}(s)\right]v(0)}{s^2 + s \gamma \bar{K}(s) + \omega^2} + \frac{\bar{\mu}(s)}{s^2 + s \gamma \bar{K}(s) + \omega^2} + \frac{\bar{\xi}(s)}{s^2 + s \gamma \bar{K}(s) + \omega^2} \;,
\end{equation}
where it was assumed $t_0=0$ and the definition of the Laplace transform operator
\begin{equation} \label{laplace_definition}
\mathcal{L}\left\{ f\right\}(s) = 
\bar{f}(s) = \int_{0}^{\infty} dt \exp{(-st)} f(t) \;, 
\end{equation}
was used. The formal solution for the time evolution of the membrane potential is then given by:
\begin{equation}
v(t) = \mathcal{L}^{-1}\left\{ \bar{v}\right\}(t) =   
\varphi(t) + \int_{0}^{t} dt' h(t-t') \mu(t') + \int_{0}^{t} dt' h(t-t') \xi(t')\;,
\end{equation}
where we have again resorted to the convolution theorem and defined
\begin{equation}
\varphi(t) = \mathcal{L}^{-1} \left\{  \frac{\dot{v}(0)+\left[ s + \gamma  \bar{K}(s)\right]v(0)}{s^2 + s \gamma \bar{K}(s) + \omega^2}     \right\}(t)\;, 
\end{equation}
as well as
\begin{equation}
h(t) =  \mathcal{L}^{-1}\left\{ \frac{1}{s^2 + s \gamma \bar{K}(s) + \omega^2}     \right\}\ (t);.
\end{equation}

Noting that the noise term $\xi$ is of zero mean, we can obtain the average dynamics of the potential $v(t)$:
\begin{equation} \label{average_v}
\langle v(t) \rangle = \varphi(t) +  \int_{0}^{\infty} dt'h(t-t') \mu(t')\;.
\end{equation}

For the case of a constant $\mu$, Eq.~(\ref{average_v}) reduces to
\begin{equation} \label{constant_mu}
\langle v(t) \rangle = \varphi(t) + \mu \mathcal{L}^{-1}\left\{ \frac{1/s}{s^2 + s \gamma \bar{K}(s) + \omega^2}     \right\}(t)\;.
\end{equation}

Once the form of the memory kernel $K(t-t')$ is specified, we can easily evaluate Eq.~(\ref{constant_mu}) numerically or algebraically. For example, if we assume a Gamma kernel (which has as a particular case the exponentially decaying kernel) \cite{smith2011introduction,Gupta2011} or an exponentially damped harmonic kernel \cite{schimansky1990harmonic,luczka2005non}, analytical solutions are possible \cite{adelman1976generalized,Farias2009}. Since they involve very long expressions, we refrain from writing them explicitly here. The standard deviation $\sigma_v = \sqrt{\langle v^2  \rangle - \langle v  \rangle^2}$ of the membrane potential can be straightforward obtained considering Eq.~(\ref{constant_mu}) and by noting that  
\begin{equation} \label{v2}
\langle v(t)^2 \rangle = \varphi(t)^2 + 2\varphi(t)\int_{0}^{t}dt'h(t-t')\mu(t')+\left[\int_{0}^{t}dt'h(t-t')\mu(t')\right]^2 + \int_{0}^{t} dt'' h(t-t'') \int_{0}^{t} dt' h(t-t') \langle \xi(t') \xi(t'')\rangle\;. 
\end{equation}

Like in the first moment case, the second moment given by Eq.~(\ref{v2})  can also be analytically obtained if, additionally to the memory kernel, a suitable correlation function for the noise term is provided.

We point out that Karmeshu and Kadambari (Ref.~\cite{Gupta2011}) proposed a different approach  to introduce memory effects and also obtained interesting results. They generalized the standard first-order SLIF model by introducing a distributed delay in the following way:
\begin{equation} \label{karmeshu}
\frac{dv}{dt} = -\beta \int_{0}^{t} dt' K(t-t') v(t') + \mu + \sqrt{2\sigma^2} \eta(t) \;, 
\end{equation}
where $\eta(t)$ has the properties of a Gaussian white noise.

\section{Numerical approach and spike train statistics} \label{numerical}

\subsection{Numerical scheme}

One of the main sources of information about the dynamics of a single neuron or an ensemble of neurons is the ISI distribution. The problem of obtaining such a distribution is formally known as a first passage time (FPT) problem \cite{van1992stochastic,hanggi1990reaction}. Despite the indisputable success of the standard formulations  on obtaining closed or approximated expressions for the FPT in the most diverse systems, the corresponding tools are not suitable to describe systems whose relaxation time is larger than the typical FPT \cite{Verechtchaguina2006, verechtchaguina2006first}. This is precisely the case of the present model and also other models that are able to produce subthreshold oscillations.

In Refs.~\cite{Verechtchaguina2006, verechtchaguina2006first}, Verechtchaguina et al. considered a resonant-and-fire model subjected to colored noise and found a closed form solution for the first term of the mean FPT equation, which in turn is formulated in terms of a sum of integrals. Despite the breakthrough developments of Verechtchaguina et al.,  to improve the reliability of the approximation more terms need to be calculated semi-analytically, and then the computational cost for obtaining the approximated result rapidly approaches the one for simulating the dynamical equation. This is no less the case of the model introduced in this paper which, besides colored noise, also considers the presence of memory. For this reason, here we choose to perform an extensive numerical analysis of the problem.

There are several methods in the literature to address
integro-differential equations like Eq.~(\ref{GLE})
\cite{adelman1976generalized,bao,Farias2009a}. For the memory kernel and noise correlation we use here, it is enough to adopt the method described in Ref.~\cite{Farias2009a}. The objective is to map the nonlocal Langevin equation into a system of local stochastic differential equations. For
this purpose, we introduce the auxiliary variable $W$, defined by
\begin{equation}  \label{auxW}
W(t)=-\int_{t_0}^{t} dt^{\prime } K(t-t^{\prime })v'\left( t^{\prime }\right)
 \;. 
\end{equation}

We consider the simplest case for the memory kernel, a decaying exponential function also known as the Ornstein-Uhlenbeck kernel:
\begin{equation} \label{kernel}
K(t-t^{\prime })=\Gamma \exp{\left[-\Gamma \left|t-t^{\prime }\right|\right]}\;.  
\end{equation}

For the noise term, we impose that its time evolution obey the Ornstein-Uhlenbeck stochastic differential equation:
\begin{equation} \label{noiseesta}
\dot{\xi}\left( t\right) =-\Gamma_{\xi} \xi \left( t\right) -\sqrt{2\Gamma_{\xi}^2\sigma_{\xi}^2}\;\eta_{\xi}\left(
t\right) \;, 
\end{equation}
where $\eta_{\xi}\left( t\right)$ is a white Gaussian noise
satisfying
\begin{eqnarray} \label{whitexizeta}
\left\langle \eta_{\xi} \left( t\right) \right\rangle  &=&0\;, \nonumber \\
\left\langle \eta_{\xi}\left( t\right) \eta_{\xi} \left( t^{\prime }\right) \right\rangle  &=& \delta (t-t^{\prime })\;.
\end{eqnarray}

We use the stationary solution of Eq.~(\ref{noiseesta}) which, together with Eq.~(\ref{whitexizeta}), yields the following correlation function for the noise:
\begin{equation} \label{colorxizeta}
\left\langle \xi\left( t\right) \xi\left( t^{\prime }\right) \right\rangle  = \sigma_{\xi}^2 \Gamma_{\xi} \exp{\left[-\Gamma_{\xi} \left|t-t^{\prime }\right|\right]} \;. 
\end{equation}
Now we can use the definition Eq.~(\ref{auxW}) and the differential equation Eq.~(\ref{noiseesta}) in order to rewrite the integro-differential Eq.~(\ref{GLE}) in terms of the following system of first order local differential equations:
\begin{align} \label{sistema}
\dot{v}\left( t\right)  =&y\left( t\right)    \notag \\
\dot{y}\left( t\right)  =&\mu-\omega^2 v +\gamma W(t) + \xi(t)\nonumber \\ 
\dot{W}(t) =&-\Gamma W(t) - K\left( 0\right) y\left( t\right)   \notag \\
\dot{\xi}\left( t\right) =&-\Gamma_{\xi} \xi \left( t\right) -\sqrt{2\Gamma_{\xi}^2\sigma_{\xi}^2}\;\eta_{\xi}\left(
t\right)  \;. 
\end{align}

The numerical model is completed upon a specification of the fire-and-reset rule. Here it reads: if $v(t)=v_{\rm{th}}$ then (i) a spike is considered to have occured at time $t$ and (ii) $v \rightarrow v_{\rm{r}}$, $y \rightarrow 0$, and $W \rightarrow 0$.

We have used a standard fourth-order Runge-Kutta method with a time step size varying between $\Delta t=10^{-2}$ and $\Delta t = 10^{-3}$ to obtain the numerical results shown in the following. This numerical scheme was found to be enough both for numerical stability and numerical precision.

\subsection{Standard measuring tools}

Neuronal dynamics can be characterized by several quantities \cite{koch2004biophysics,Vilela2009a,Vilela2009,Burkitt2006} and here we define some of them that will be useful in the analysis we perform in the next sections. The \textit{spike time} is defined as the time instant when the membrane potential reaches the threshold value. Given a time window $\Delta t_{W} = t_f - t_0$ and a trial (say, the $i$th-trial), a neuron produces a finite set of spikes, whose temporal distribution can be accounted for by
\begin{equation} \label{spiketrain}
y_i(t) = \sum_{j=1}^{N_i} \delta(t-t_j^{i})\;,
\end{equation}
with $t_1^{i} > t_0$ and $t_{N_i}^{i} < t_f$. The distribution $y_i(t)$ is called \textit{spike train}, and the sum runs discretely over each spike time $t_j^{i}$, from $j=1$ to $j=N_i$. A consistency relation between $y_i(t)$ and $N_i$ can be written. The number of spikes in a given trial can be evaluated as
\begin{equation} \label{numberspikes}
N_i = \int_{t_0}^{t_f} dt y_i(t) \;. 
\end{equation}
In $n$ trials, the total number of spikes is defined as $\mathcal{N} = \sum_{i=1}^n N_i$. 
We define the firing rate as an average over time:
\begin{equation} \label{firingrate}
r =  \frac{\mathcal{N}}{n\Delta t_W}\;.
\end{equation}

The time interval between two successive spikes ( $ T_j^{i} = t_{j+1}^{i} - t_{j}^{i}$) is called the \textit{interspike interval} (ISI). This definition allows us to concatenate the ISI's of all $n$ trials and then define the following quantities: the average ISI (denoted by $\langle T \rangle$), 
\begin{equation}
\langle T \rangle = \frac{1}{\mathcal{N}-n} \sum_{i=1}^{n} \sum_{j=1}^{N_i-1} T_j^i\;, \label{rate}
\end{equation}
and the \textit{coefficient of variation} ($\rm{CV}$), denoted by $c_v$ in the equations:
\begin{equation}
c_v = \frac{\left[ \langle T^2 \rangle -  \langle T \rangle^2\right]^{1/2}}{ \langle T \rangle }\;, \label{CV}
\end{equation}
where $\langle T^2 \rangle$ is defined analogously to Eq.(\ref{rate}). The CV is of course a measure of the regularity of the spike train. From Eq.~(\ref{CV}), one can infer that a time uniform spike train leads to $c_v=0$, while a Poissonian spike train is characterized by $c_v=1$.
\begin{figure}[htb]
 \centerline{
   \psfig{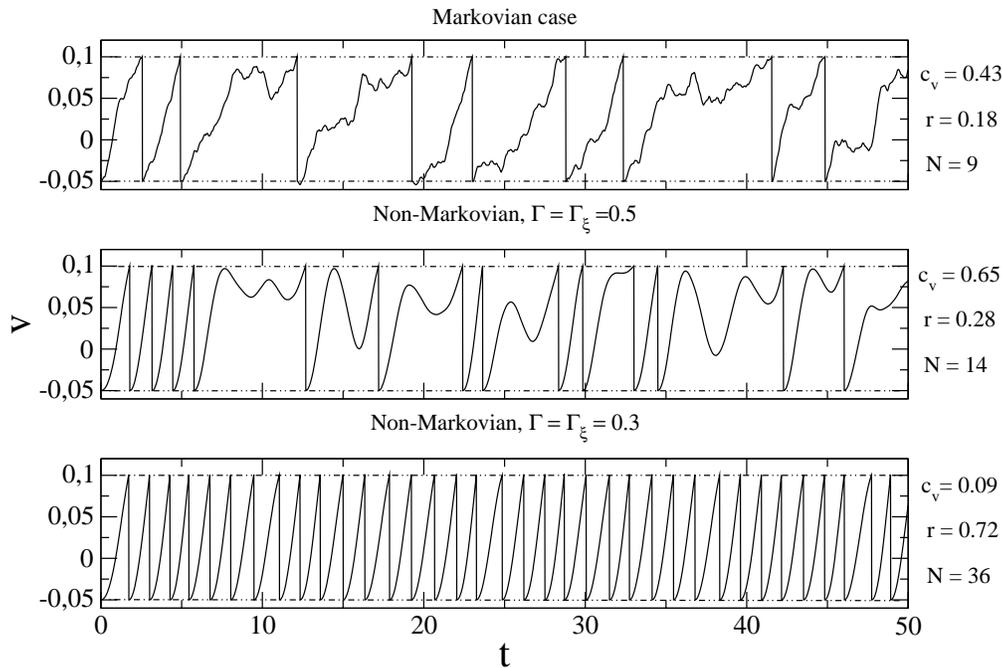}
   }
   \caption{\sf Sample trials of the potential $v(t)$ obtained from Eq.(\ref{sistema}) in the tonic spiking regime using $\sigma_{\xi}=0.1$. }
   \label{single}
\end{figure}
%


\section{Results and discussions}
\label{results}

In the following numerical simulations, we have considered statistics based on at least $n=10^4$ trials where each trial is characterized by an observational time window $\Delta t_W = 10^3$. These choices were made in order to obtain a sample space composed by at least $10^6$ spikes. In all cases, we set a threshold value $v_{\rm{th}} = 0.1$ and a reset value $v_r=-0.05$. We distinguish between two characteristic regimes: in subsection \ref{det} we analyze the neuron dynamics when $\mu/\omega^2> v_{\rm{th}}$ (tonic spiking regime, $\mu= 0.2$), and in subsection \ref{noisedriven} we analyze the situation where $\mu/\omega^2< v_{\rm{th}}$, the so-called noise-induced spiking regime ($\mu= 0.08$).  We considered in all cases $\gamma=5$ and $\omega=1$, corresponding to the overdamped case. Therefore, when present, the subthreshold oscillations are due to the presence of memory.

\subsection{Tonic spiking regime} \label{det}

We start by illustrating the dynamics provided by Eq.~(\ref{sistema}) showing in Fig.~\ref{single} some sample trials in a reduced time window for different memory parameter $\Gamma$. 
\begin{figure}
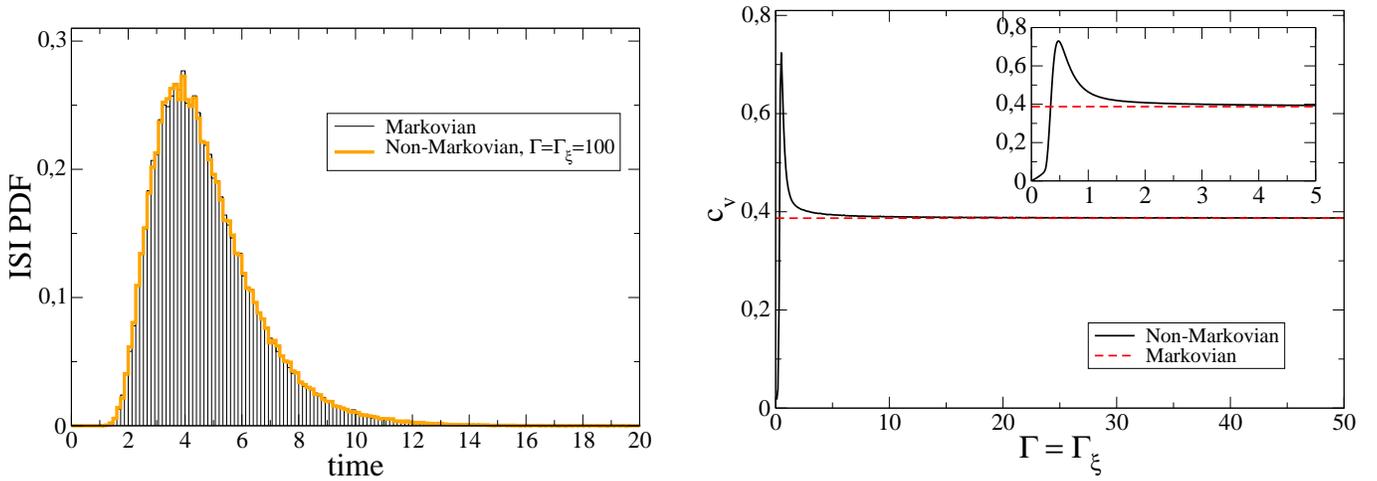

 \centerline{\psfig{file=fig2a.eps,scale=0.35,angle=0} \hspace{0.5cm}
\psfig{file=fig2b.eps,scale=0.35,angle=0}
   }
   \caption{(Color online) Convergence to the Markovian dynamics. \sf Left: probability density function for the interspike interval obtained using Eq.(\ref{sistema}) and its Markovian approximation. Right: the coefficient of variation (CV) as a function of $\Gamma$. The inset plot shows a detailed view of the CV for shorter values of $\Gamma$. Note the nonmonotonic behavior of CV as we increase $\Gamma$ (Markovian case: $\Gamma=\Gamma_{\xi}\rightarrow \infty$). In both plots, $\sigma_{\xi}=0.1$ and $\mu=0.2$.}
   \label{marknonmark_histogram_CV}
\end{figure}

In Fig.~\ref{marknonmark_histogram_CV}, we test the limiting behavior of our numerical code. As expected, for large values of $\Gamma$ and $\Gamma_{\xi}$, we observe a good agreement between the non-Markovian dynamics and the Markovian approximation, given by Eq.~(\ref{RIF}) with $\lambda=1$: on the left panel of Fig.~\ref{marknonmark_histogram_CV} we show this agreement for the probability density function of the interspike intervals. On the right panel, we show the approach of the coefficient of variation to its counterpart in the Markovian limit for large values of $\Gamma$.

\begin{figure}
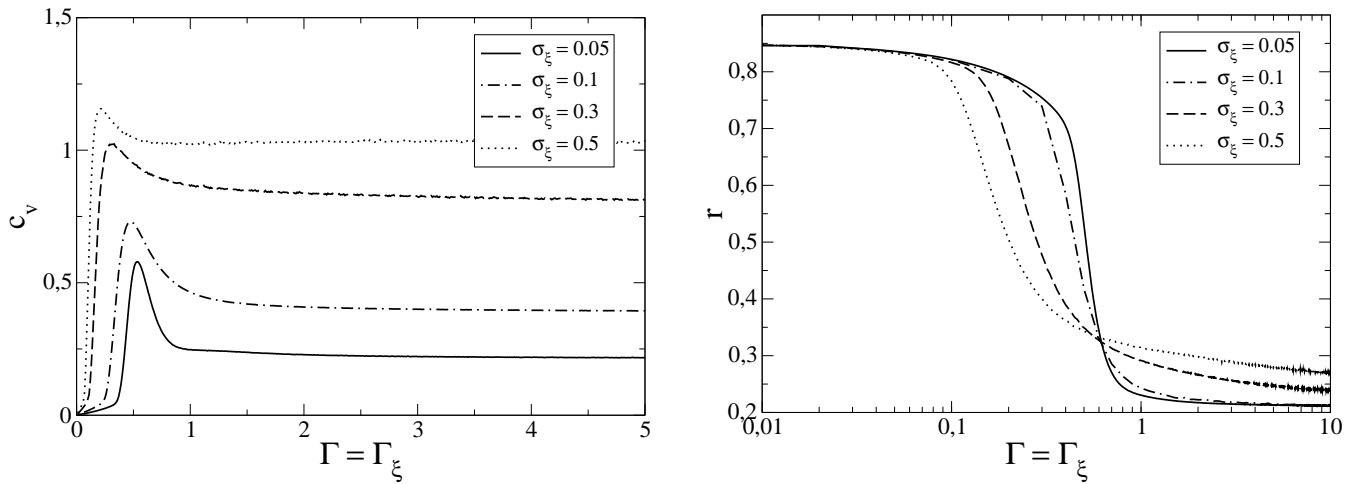

\centerline{ \psfig{file=fig3a.eps,scale=0.35} \hspace{0.5cm}
\psfig{file=fig3b.eps,scale=0.35}
  }
\caption{The behavior of the $\rm{CV}$ and the firing rate $r$ as a function of 
$\Gamma=\Gamma_{\xi}$ in the tonic spiking regime ($\mu=0.2$).}
\label{gamma}
\end{figure}

In the inset plot of Fig.~\ref{marknonmark_histogram_CV}, we call attention to the non-monotonic behavior of the non-Markovian CV as a function of the memory parameter. In the long memory limit ($\Gamma = \Gamma_{\xi} \ll 1$), the CV is near zero, what characterizes a very regular spike train. If we proceed increasing $\Gamma$, the value of CV also increases until a local maximum is reached ($\Gamma = \Gamma_{\xi} \approx 0.5$). After that, the value of CV starts to decrease until it reaches the Markovian limit ($\Gamma = \Gamma_{\xi} \gg 1$). In Fig.~\ref{gamma} (left panel) we observe that increasing the noise intensity shifts the  CV curves toward a higher level of variability, and also causes a softening effect on the shape of the CV peak. On the right panel of the same figure, we show the impact of noise on the mean firing rate $r$: increasing $\sigma_{\xi}$ results in an appreciable decrease of the firing rate in the approximated range $0.1 < \Gamma = \Gamma_{\xi} < 0.8$, and in an increase of the firing rate for $\Gamma = \Gamma_{\xi} > 0.8$, approximately.

The non-trivial behavior induced by memory described above, which is also present in the noise-induced spike regime (Sec.IV B), can be seen as an effective representation of some underlying mechanism that acts controlling the firing variability and rate of a given neuron. A number of such mechanisms have been reported in the literature. 
For instance, it has been suggested that GABAergic autaptic transmission (GAT) is responsible for the regulation of spike-timing precision in the so-called \textit{fast spiking interneurons}, leading to the suppression of variability \cite{bacci2006enhancement,bacci2003functional}. In the case of our model, the memory is introduced to reflect a property of the input. This has a strong analogy with the experimental results reported in Refs.~\cite{bernander1991synaptic} and \cite{fernandez2009reduction}, which show 
that the overall activity level of the network that embeds a given neuron  modifies the integrative properties and the performance of that single cell.

\begin{figure}[htb!]
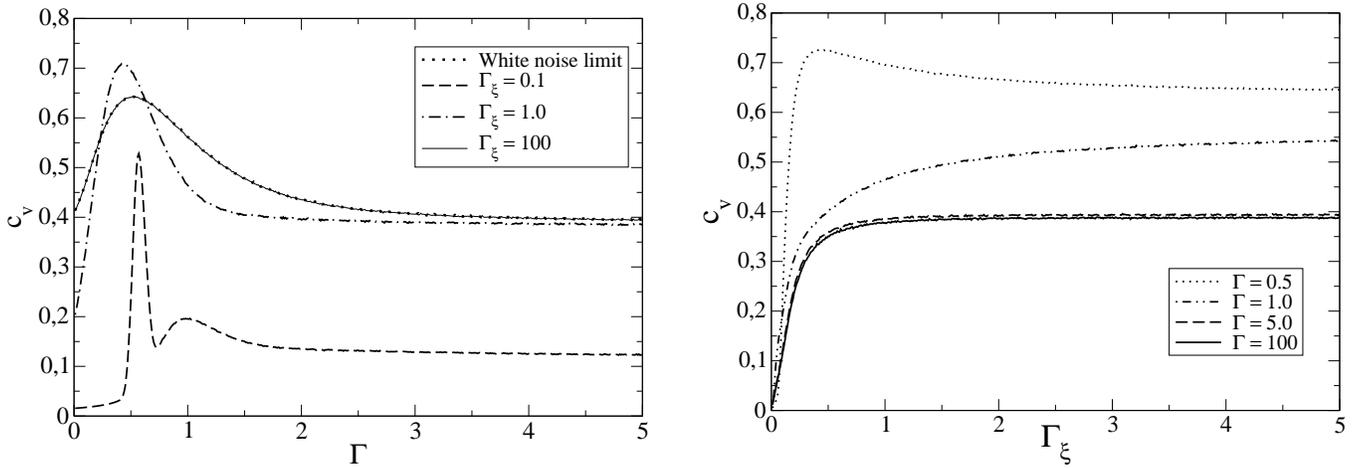

\centerline{ \psfig{file=fig4a.eps,scale=0.35} \hspace{0.5cm}
\psfig{file=fig4b.eps,scale=0.35}
  }
\caption{Dissociated effects of memory and colored noise. Left: $\rm{CV}$ as a  function of $\Gamma$ for different values of $\Gamma_{\xi}$. Right: Suppression of the peak in the memoryless limit, $\Gamma \gg 1$. In both panels, $\mu=0.2$ and $\sigma_{\xi}=0.1$.}
\label{whitevscolored}
\end{figure}

It is important to note that it is the presence of memory rather than that of colored noise the determinant factor that induces the emergence of a peak in the  $c_v \times$ $\Gamma$ curve. If we ease the restriction $\Gamma=\Gamma_{\xi}$ (used only to reduce the number of free parameters) and consider Eq.~(\ref{sistema}) in the white noise limit ($\Gamma_{\xi} \rightarrow \infty$), we obtain the result shown in the left panel of Fig.~\ref{whitevscolored}: the dotted line, that essentially coincides with the $\Gamma_{\xi}=100$ curve (full line), still exhibits the local maximum of the CV as a function of $\Gamma$.  Note that, in sharp contrast, no maximum  occurs for the CV as a function of $\Gamma_{\xi}$ when we consider the memoryless limit ($\Gamma \rightarrow \infty$) of the dynamical equation (approximated by the $\Gamma=100$ curve on the right panel of Fig.~\ref{whitevscolored}). Looking again at the left panel of Fig.~\ref{whitevscolored}, one notes that the value of $\Gamma_{\xi}$ determines how smooth the peak structure is. The smaller the value of $\Gamma_{\xi}$ is, the sharper is the peak, and even a double-peak structure can emerge in the  $c_v \times \Gamma$ curve (dashed line, $\Gamma_{\xi}=0.1$). The bigger the value of   $\Gamma_{\xi}$ is (until it saturates in the white noise limit), the smoother is the shape of the peak. Also, increasing $\Gamma_{\xi}$ implies a global upward shift in the neuron variability, situation that tends to be slightly violated  only in the vicinity of the peak when $\Gamma_{\xi} \gg 1$, since the peak becomes increasingly smooth as $\Gamma_{\xi}$ is increased.

\begin{figure}[htb!]
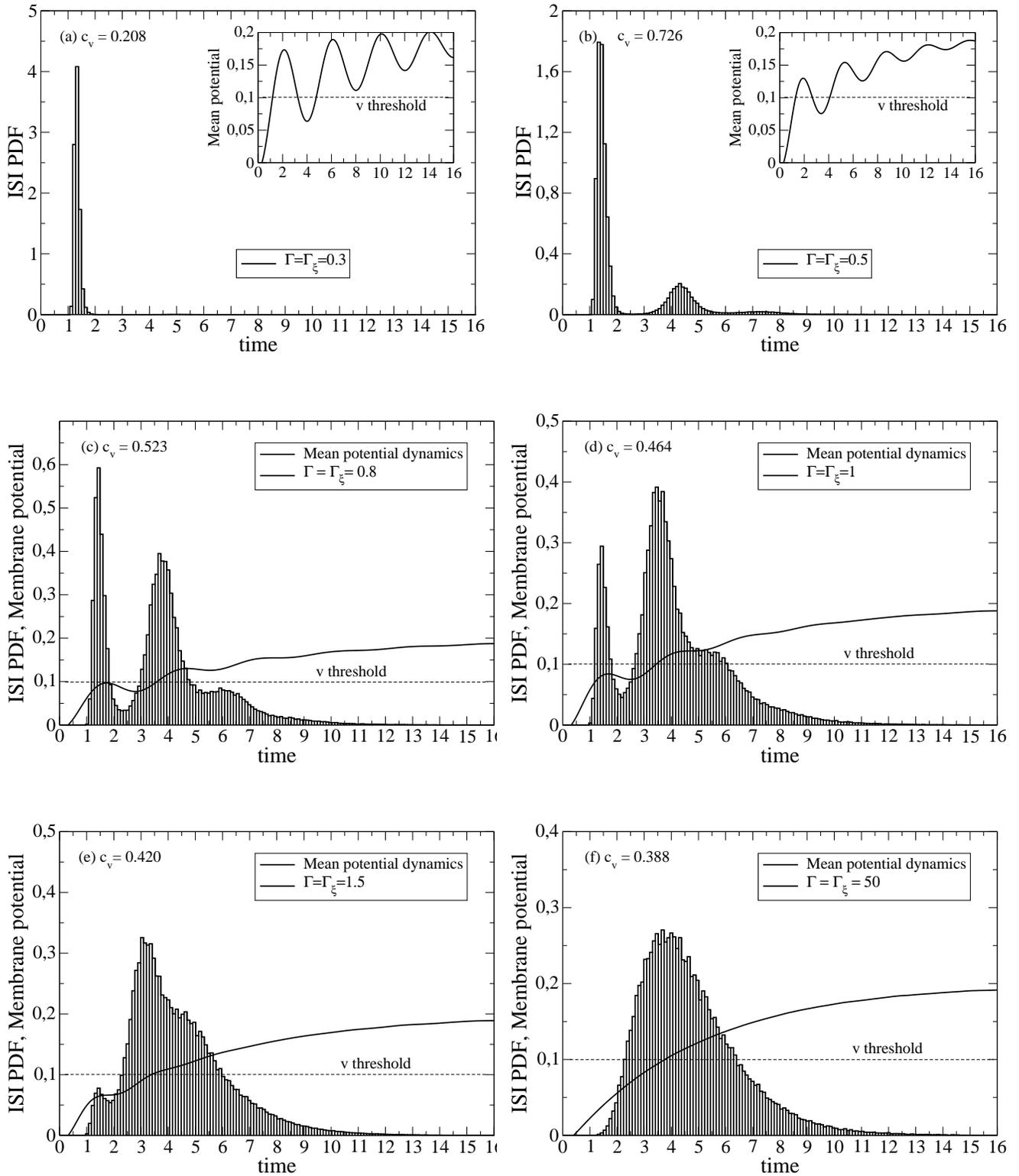

\centerline{ \psfig{file=fig5a.eps,scale=0.35,angle=0} \hspace{0.3cm}
\psfig{file=fig5b.eps,scale=0.35,angle=0}   }
\vspace{1cm}
\centerline{ \psfig{file=fig5c.eps,scale=0.35,angle=0}
\psfig{file=fig5d.eps,scale=0.35,angle=0}   }
\vspace{1cm}
\centerline{ \psfig{file=fig5e.eps,scale=0.35,angle=0}
\psfig{file=fig5f.eps,scale=0.35,angle=0}
  }
\caption{Emergence of bimodality in the ISI distribution. In all panels, $\mu=0.2$ and  $\sigma_{\xi}=0.1$.}
\label{histograms}
\end{figure}

A joint analysis of the subthreshold evolution of Eq.~(\ref{sistema}) and the ISI histograms for a given time scale $\Gamma^{-1}$ is useful to shed light on the mechanism that leads to the emergence of the peak in the $ c_v \times \Gamma$ curve. In Fig.~\ref{histograms} we take a closer look at the case  $\sigma_{\xi}=0.1$  by showing a sequence of snapshots for fixed values of $\Gamma=\Gamma_{\xi}$. In panel (a), we set $\Gamma=\Gamma_{\xi}=0.3$, a value to the left of the maximum $\Gamma=\Gamma_{\xi} \approx 0.5$ (see Fig.~\ref{gamma}), and it can be noted that the ISI distribution exhibits a very sharp unimodal structure, which is characterized by a small value of CV, as shown in Fig.~\ref{gamma}. These findings can be easily understood if one looks at the inset plot, which shows the mean membrane potential trajectory (Eq.~\ref{constant_mu}) crossing the threshold value at a very early time instant due to memory-induced oscillations. The amplitude of these oscillations is so large that non-crossing trajectories become unlikely already after the first peak of the mean potential. When $\Gamma=\Gamma_{\xi}=0.5$ (panel (b)), we can note a decrease in the height of the first ISI peak and the emergence of a second one. This configuration of the ISI distribution characterizes the maximum CV, since trajectories that are able to avoid the threshold crossing at the first oscillation do not possess a significantly dominant most probable interspike interval, since the emergent mode does not exhibit a sufficiently pronounced peak. Setting $\Gamma=\Gamma_{\xi}=0.8$ (panel (c)), the second mode  becomes well defined, since its height is now comparable to the one of the first peak. Thus, in this configuration we can infer a decrease in the CV value, what is corroborated by Fig.~\ref{gamma}. This decreasing behavior of the CV persists from this panel to the last one. Note that for larger values of $\Gamma=\Gamma_{\xi}$, we have a continuous suppression of the bimodality, as can be seen in panels (e) and (f). However, even though both (a) and (f) ISI distributions exhibit a single-peaked structure, they yield different values of the CV: as the system moves toward the Markovian limit ($\Gamma=\Gamma_{\xi} \gg 1$) \textit{plateau} exhibited in the right panel of Fig.~\ref{marknonmark_histogram_CV}, the CV decreases, but it saturates in a value larger than the one obtained when $\Gamma=\Gamma_{\xi} \ll 1$.

\begin{figure}[htb!]
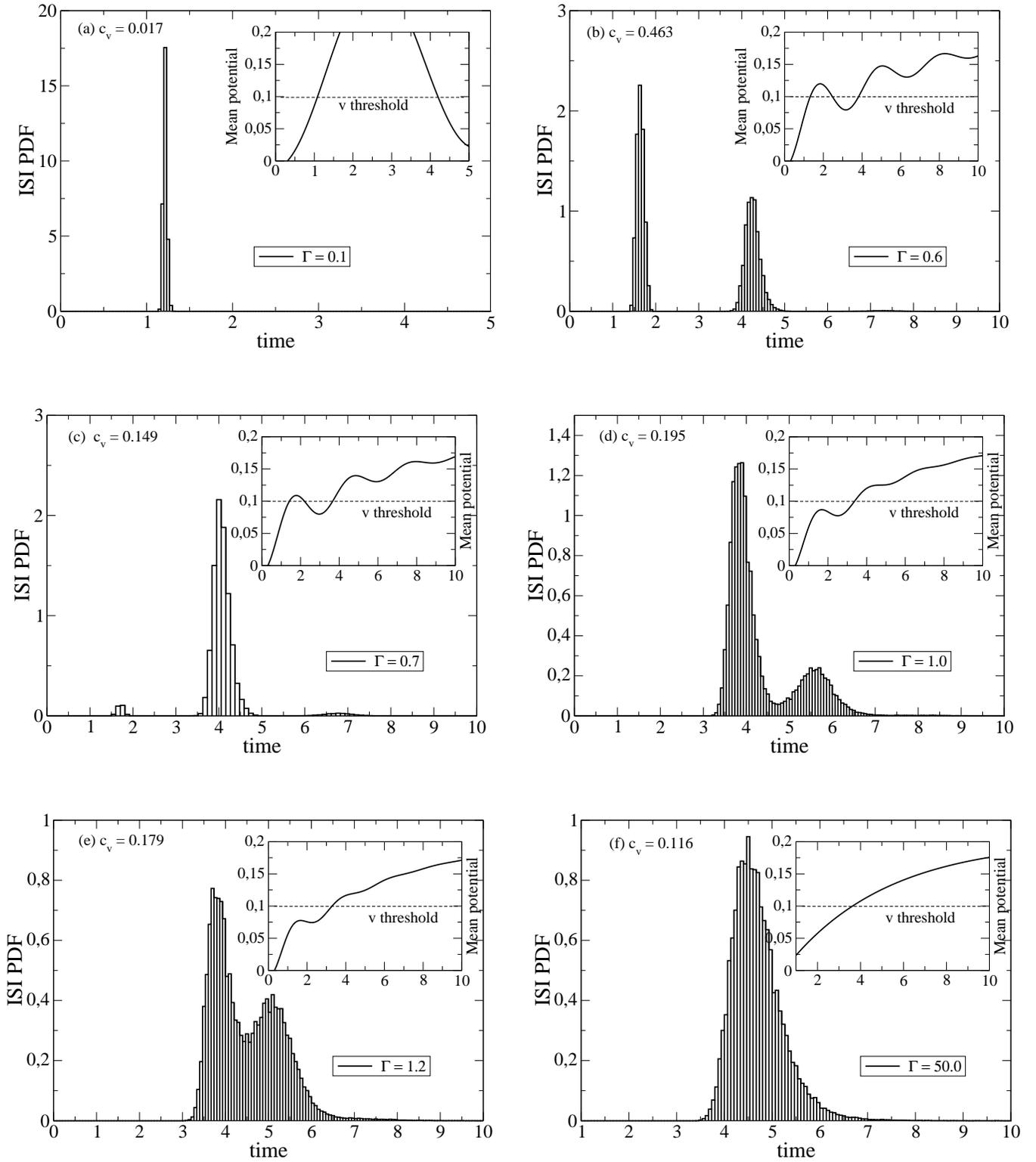

\centerline{ \psfig{file=fig6a.eps,scale=0.35,angle=0} \hspace{0.5cm}
\psfig{file=fig6b.eps,scale=0.35,angle=0}   }
\vspace{1cm}
\centerline{ \psfig{file=fig6c.eps,scale=0.35,angle=0} \hspace{0.5cm}
\psfig{file=fig6d.eps,scale=0.35,angle=0}   }
\vspace{1cm}
\centerline{ \psfig{file=fig6e.eps,scale=0.35,angle=0} \hspace{0.5cm}
\psfig{file=fig6f.eps,scale=0.35,angle=0}
  }
\caption{Emergence of bimodality in the ISI distribution: analysis of the double-peak structure in the $\Gamma_{\xi}=0.1$ case of Fig.~\ref{whitevscolored}. In all panels: $\Gamma_{\xi}=0.1$, $\mu=0.2$ and $\sigma_{\xi}=0.1$.}
\label{histograms_Gx0p1}
\end{figure}

The same analysis can be performed to clarify the double-peak structure in the $c_v \times \Gamma$ curve that is exhibited in the $\Gamma_{\xi}=0.1$ case of Fig.~\ref{whitevscolored}. In Fig.~\ref{histograms_Gx0p1} we show the ISI distribution for representatives values of $\Gamma$. From panel (a) to panel (b), we note the transition from a sharp, unimodal distribution to a  distribution with two dominant modes. Panel (b) represents the maximum CV configuration for this particular case. In panel (c), we fix $\Gamma=0.7$ (the minimum between the two peaks in Fig.~\ref{whitevscolored}) and a suppression of the first mode occurs, thus decreasing the CV value since only one dominant mode remains. The next panel corresponds to the second peak shown in Fig.~\ref{whitevscolored}), and from this panel to the last one, we note a decrease in the CV value until the memoryless limit is reached and the CV assumes an approximately constant value.

\begin{figure}[htb!]
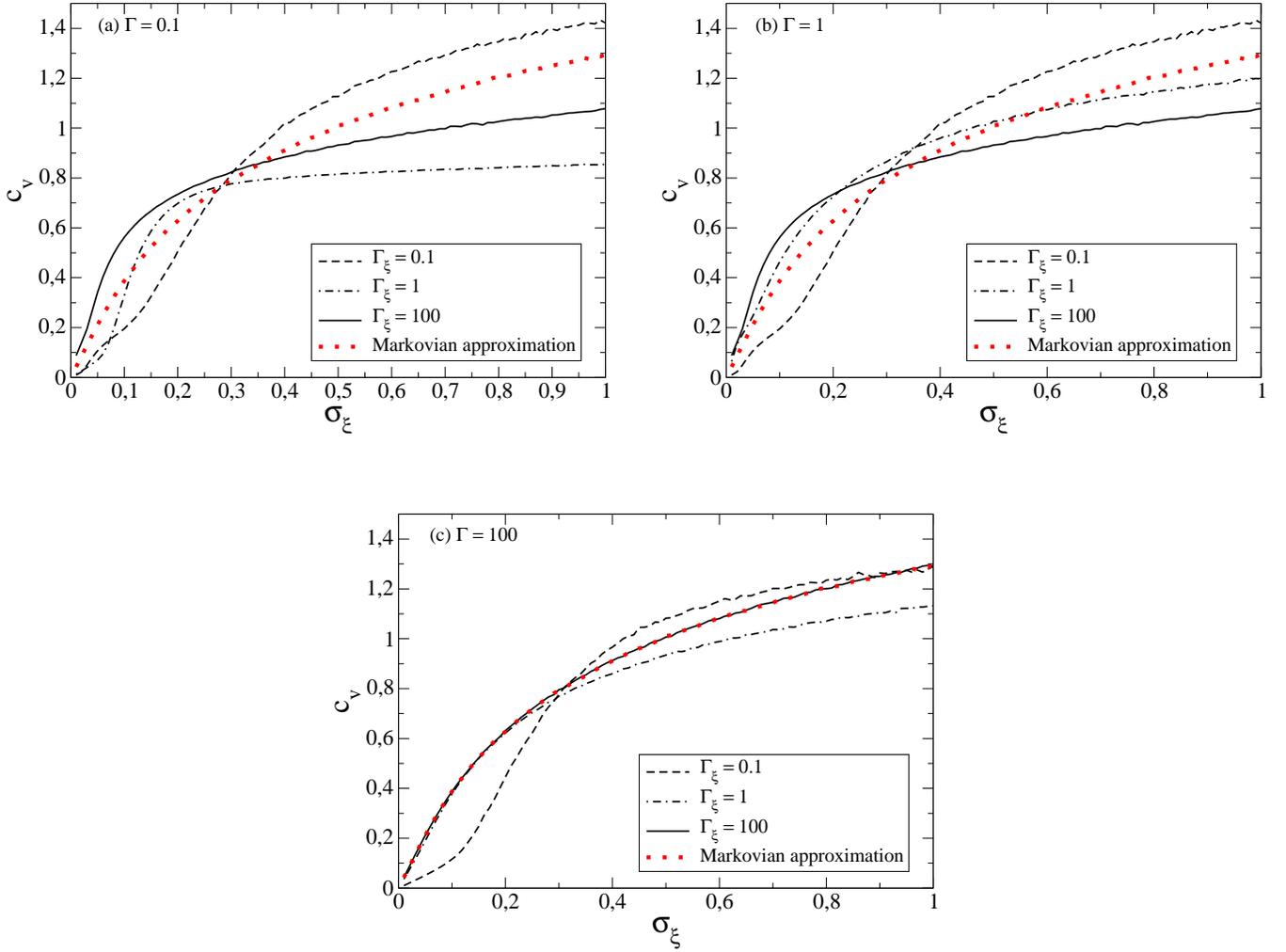

 \centerline{ \psfig{file=fig7a.eps,scale=0.35,angle=0} \hspace{0.5cm}
  \psfig{file=fig7b.eps,scale=0.35,angle=0}
}
\vspace{1cm}
\centerline{ \psfig{file=fig7c.eps,scale=0.35,angle=0}
}  
\caption{(Color online) The behavior of the CV as a function of the additive noise intensity $\sigma_{\xi}$ in the tonic spiking regime.}
\label{sx}
\end{figure}

As a last remark in this subsection, we illustrate in Fig.~\ref{sx} how the coefficient of variation behaves as a function of the noise intensity $\sigma_{\xi}$. The results are intuitive, since we can note a monotonic increase in the value of the CV as the noise intensity increases. The dotted red curves that are present in all panels correspond to the Markovian approximation ($\Gamma=\Gamma_{\xi}\to \infty$). As expected, for large values of $\Gamma = \Gamma_{\xi}$ (panel (c), $\Gamma_{\xi}=100$ case), the Markovian dynamics becomes a reliable approximation for the non-Markovian dynamics.

\subsection{Noise-induced spike regime and coherence resonance} \label{noisedriven}

Now let us turn to the analysis of the noise-induced spike regime, where the external input $\mu$ and the parameter $\omega$ are defined such that $\mu/\omega^2< v_{\rm{th}}$.

\begin{figure}[htb!]
\centerline{ \psfig{file=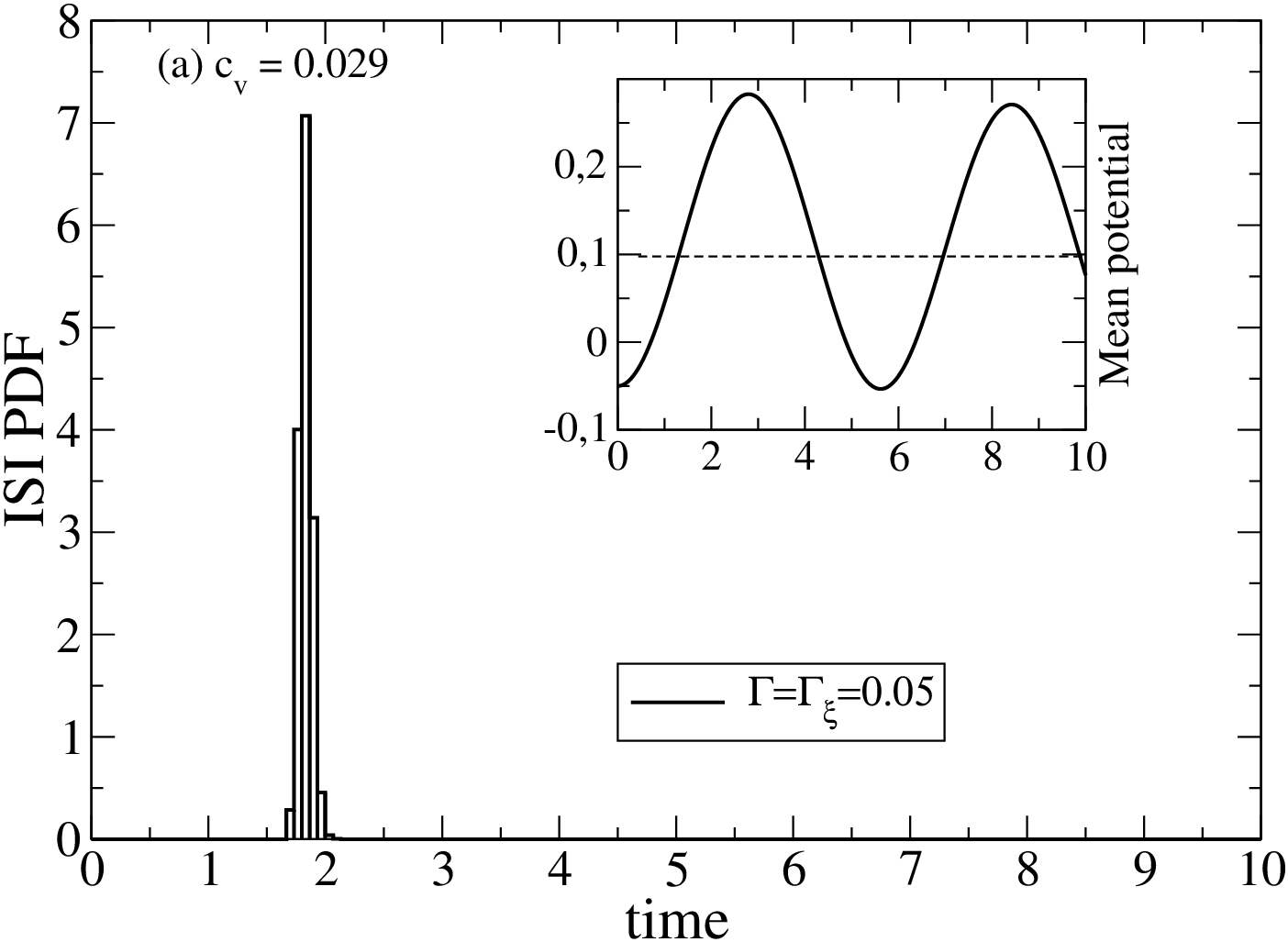,scale=0.35,angle=0} \hspace{0.5cm}
\psfig{file=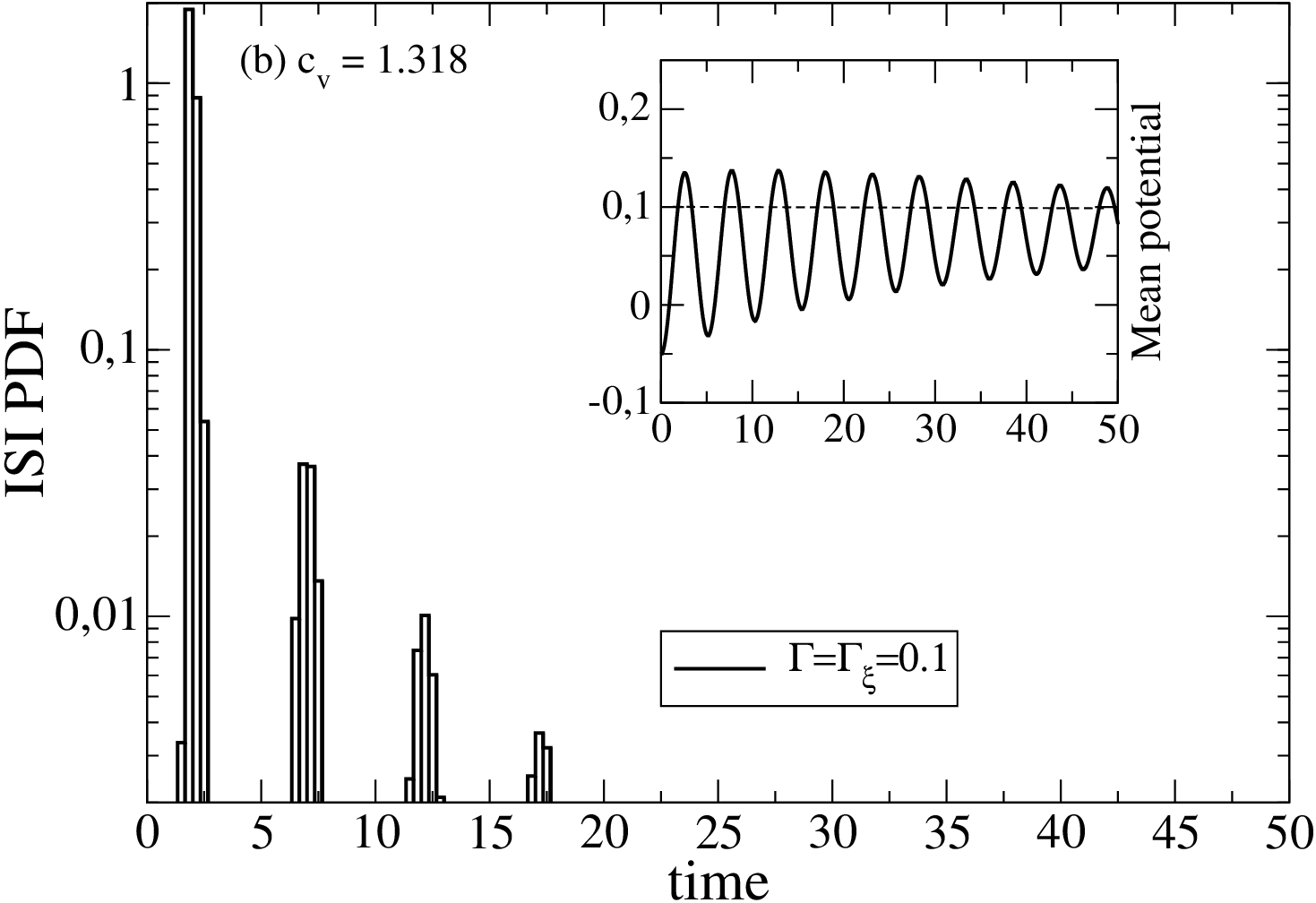,scale=0.35,angle=0}   }
\vspace{1cm}
\centerline{ \psfig{file=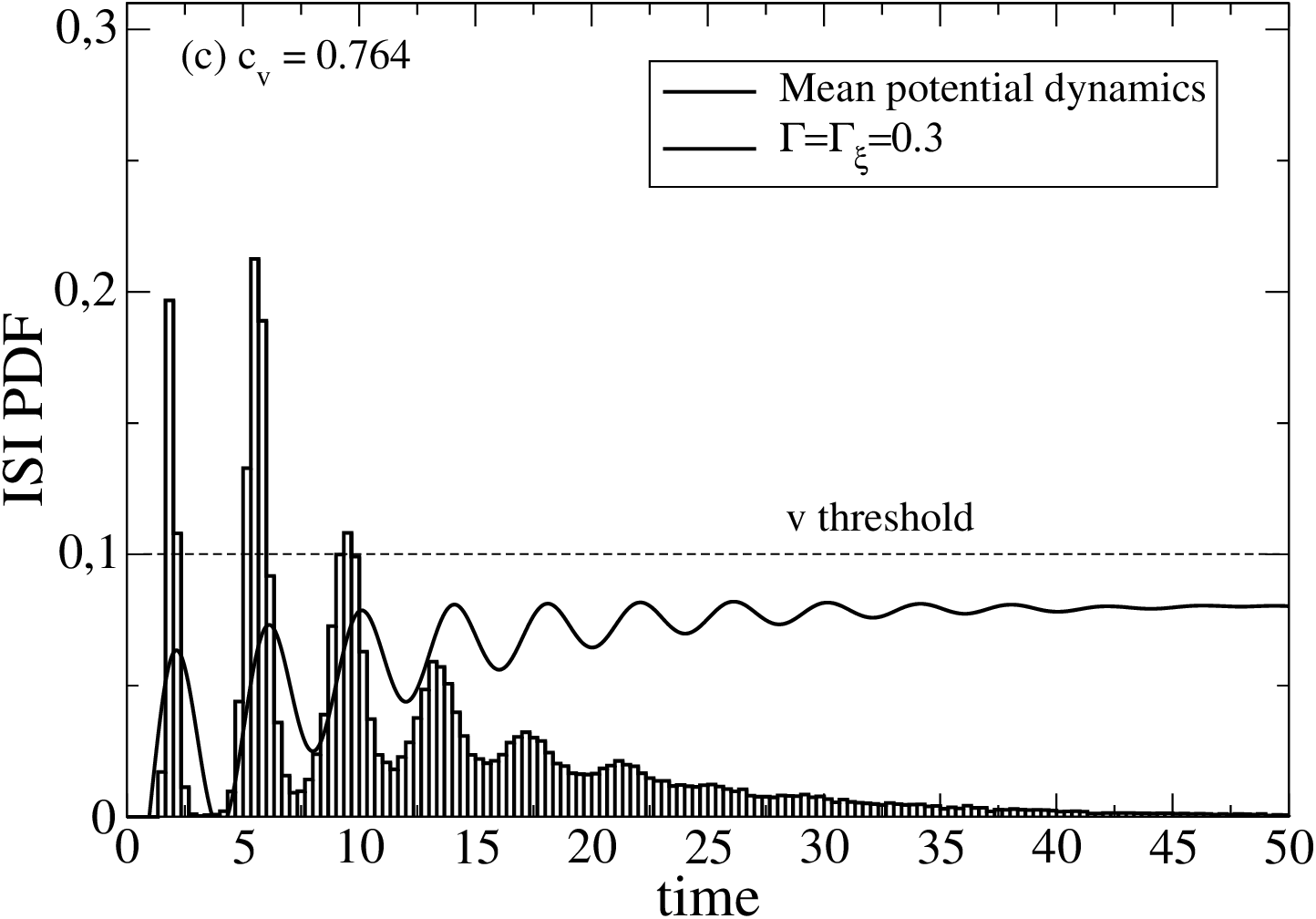,scale=0.35,angle=0} \hspace{0.5cm}
\psfig{file=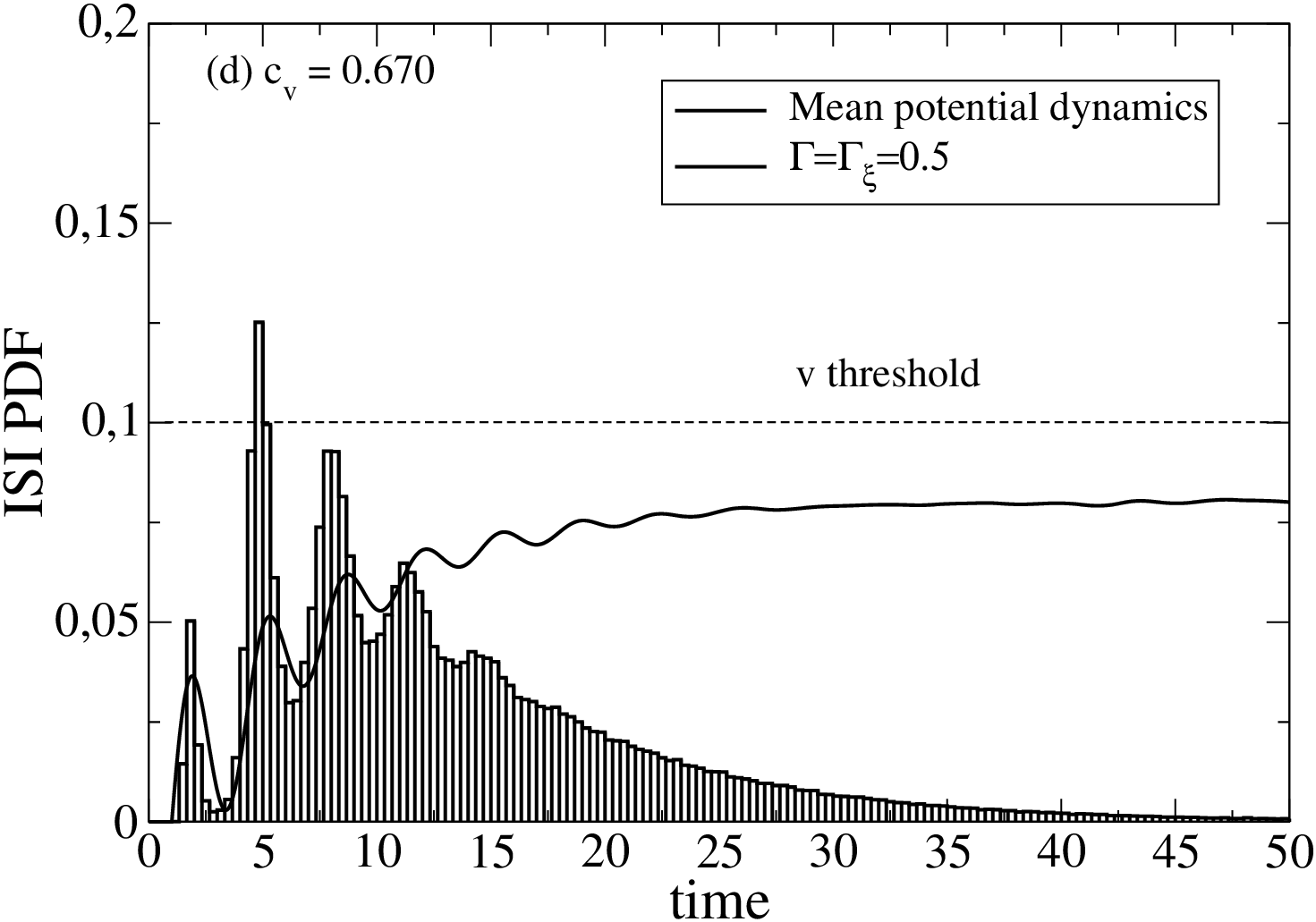,scale=0.35,angle=0}   }
\vspace{1cm}
\centerline{ \psfig{file=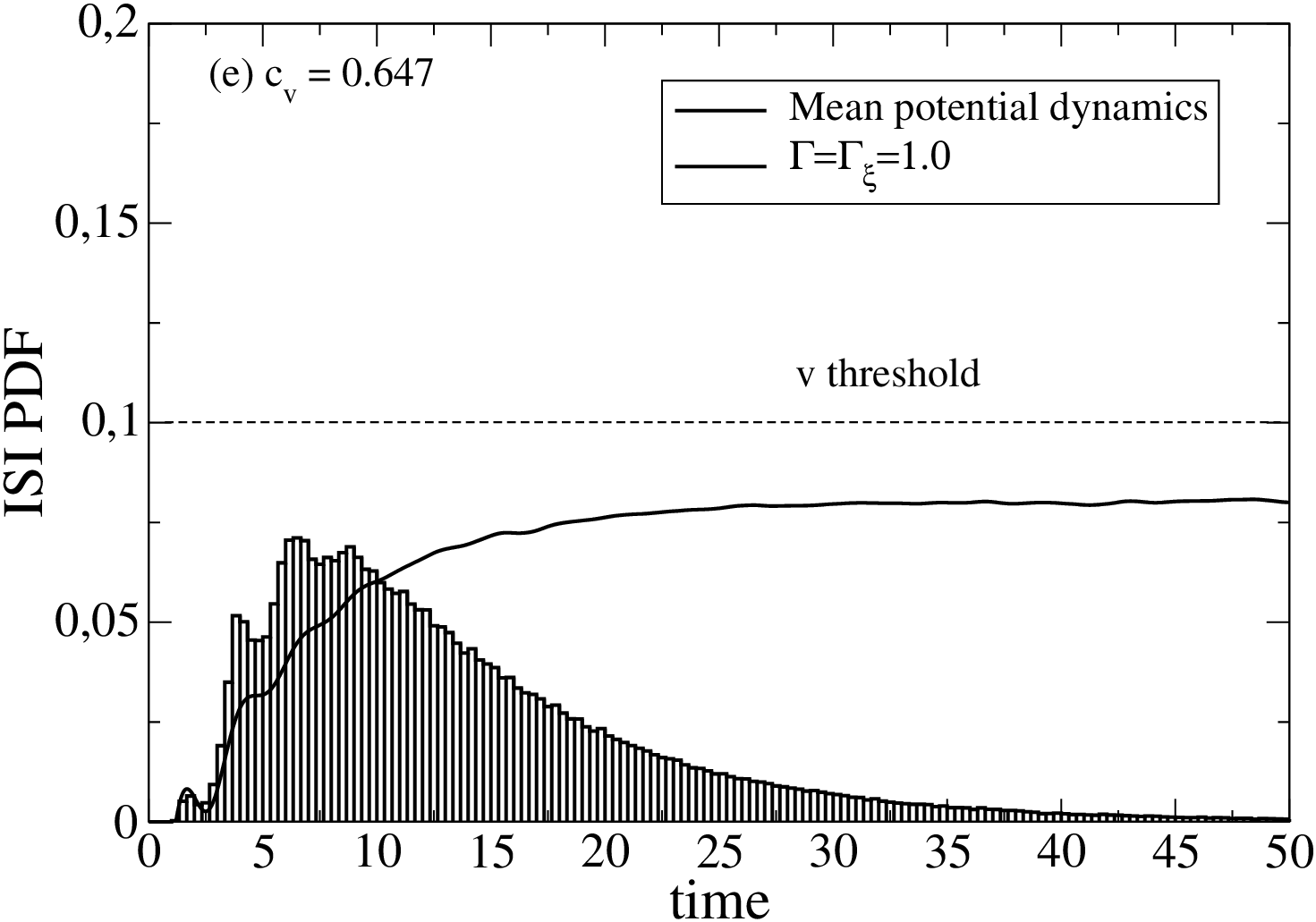,scale=0.35,angle=0} \hspace{0.5cm}
\psfig{file=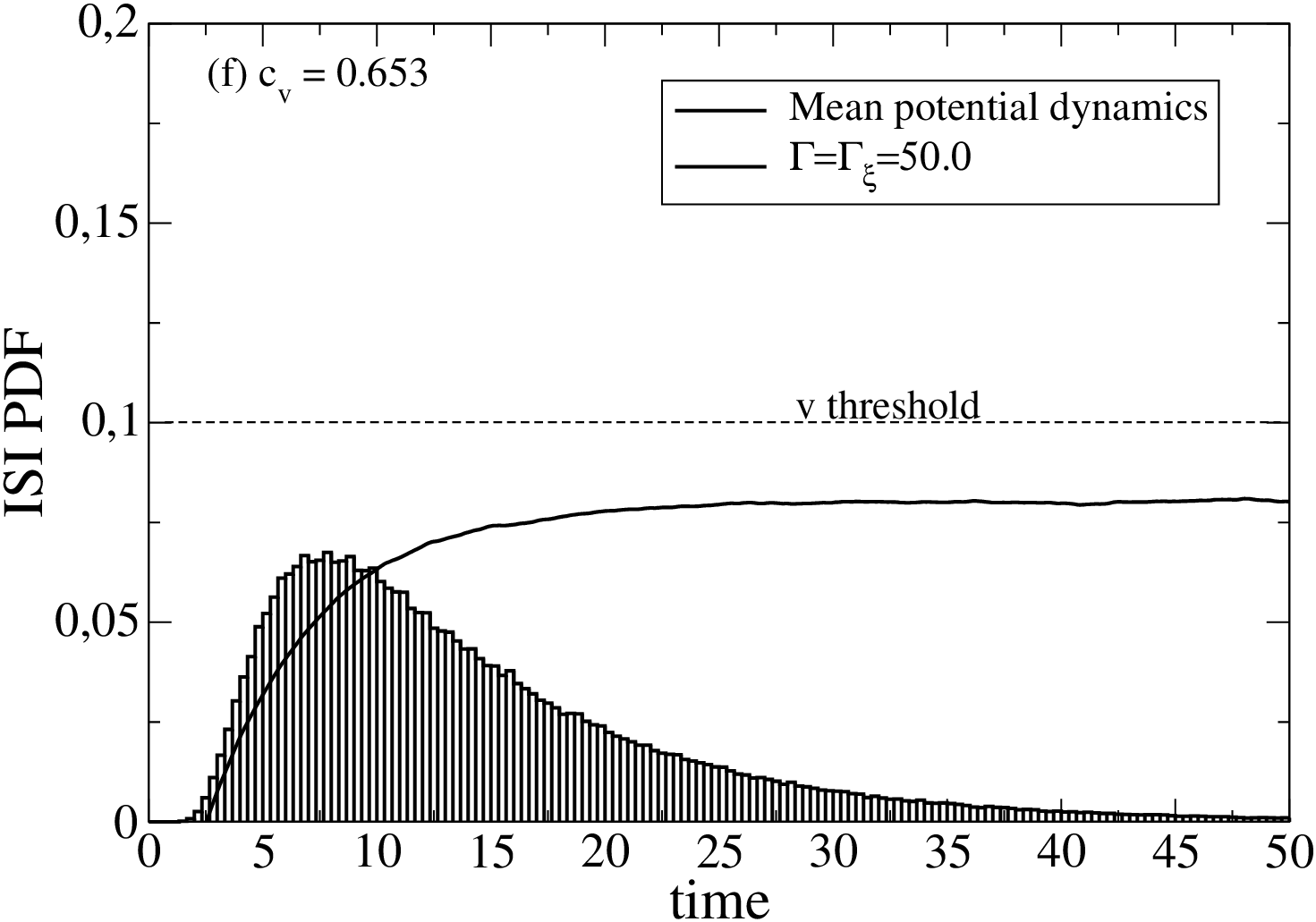,scale=0.35,angle=0}
  }
\caption{Emergence of multimodality in the ISI distribution. In all panels: $\mu=0.08$ and $\sigma_{\xi}=0.1$.}
\label{histograms_subthreshold}
\end{figure}

\begin{figure}[h!]
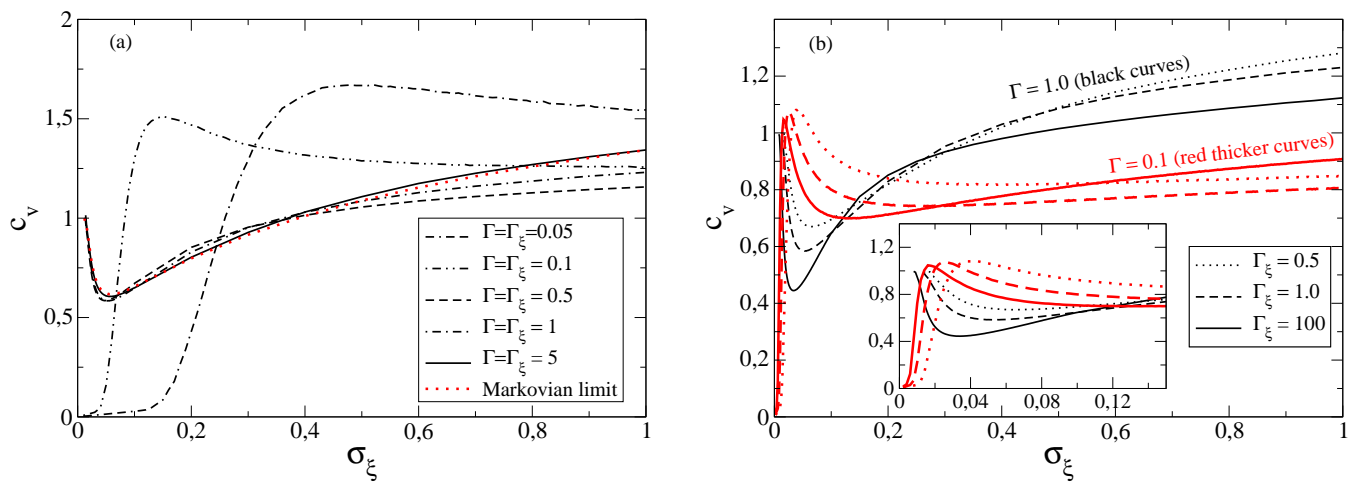

\centerline{ \psfig{file=fig10a.eps,scale=0.35} \hspace{0.5cm}
 \psfig{file=fig10b.eps,scale=0.35}
}
\caption{(Color online) The behavior of the $\rm{CV}$ as a function of the noise intensity $\sigma_{\xi}$ in the noise-induced spike regime: coherence resonance can be suppressed by memory effects.}
\label{CVsub_sx}
\end{figure}

In this regime, the qualitative behavior of both the CV and rate as functions of $(\Gamma=\Gamma_{\xi})$ is similar to their counterparts in the tonic spiking regime (Fig.~\ref{CVsub_sx}. 
However, if we take a closer look at the ISI distribution in this case (Fig.~\ref{histograms_subthreshold}, where the $\sigma_{\xi}=0.1$ case of Fig.~\ref{CVsub_sx}) is depicted), we can note the emergence of a multimodal structure that results from the memory induced subthreshold oscillations.  Like in the previous subsection,  for $\Gamma=\Gamma_{\xi}\ll 1$ (long memory limit, panel (a)) and for $\Gamma=\Gamma_{\xi}\gg 1$ (Markovian limit, panel (f)), the ISI distribution is unimodal, and for a critical value $\Gamma=\Gamma_{\xi}\approx 0.1$ (panel (b), log scale), we obtain a multimodal ISI distribution that maximizes the value of CV. It is important to note that we adopted the standard definition of the noise-induced spike regime, thus taken from the memoryless model perspective. Since memory is able to induce oscillatory behavior in the potential dynamics, it can eventually cause deterministic threshold crossing depending on the amplitude of the oscillations. For example, panels (a) and (b) show sufficiently small values of $\Gamma=\Gamma_{\xi}$ such that memory provides a transition from noise-induced to tonic spike regime. From panel (c) onward, the spike regime becomes noise-induced in the standard sense.
Also, from panel (b) to panel (e), we observe a decrease in the CV value until it reaches the Markovian limit value in panel (f). In this panel, we recover the typical unimodal ISI histogram of overdamped models.
\begin{figure}[htb!]
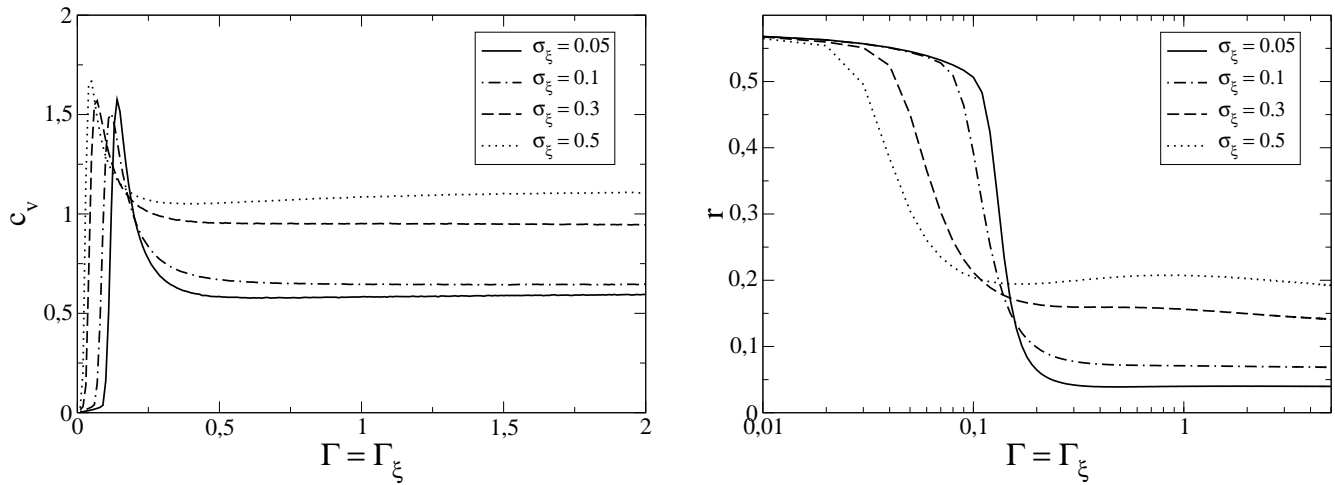

\centerline{ \psfig{file=fig8a.eps,scale=0.35} \hspace{0.5cm}
\psfig{file=fig8b.eps,scale=0.35}
  }
\vspace{0.5 cm}
\caption{The behavior of the $\rm{CV}$ and the firing rate $r$ as a function of the
$\Gamma=\Gamma_{\xi}$ in the noise induced spike regime.}
\label{gamma_sub}
\end{figure}

Now we discuss an interesting phenomenon known as \textit{coherence resonance} (CR) \cite{pikovsky1997coherence}, that can be present not only in neuronal systems \cite{lindner2002maximizing,lindner2000coherence}, but also in optical systems, plasma discharge dynamics and others \cite{lee1998coherence,giacomelli2000experimental,shaw2014intrinsic}. Like in \textit{stochastic resonance} (SR) \cite{Gammaitoni1998,Hanggi2002}, CR can be interpreted as a non-deleterious behavior induced by the presence of noise. But unlike SR, where setting an appropriate amount of noise can improve the ability of neurons detect external subthreshold signals (e.g. periodic inputs), the constructive role of noise in CR phenomenon is characterized by the emergence of order under the presence of a stochastic source only. Specifically in the neuronal dynamics case, a critical level of noise intensity can maximize the regularity (i.e. the coefficient of variation reaches a minimum) of a spike train. 
We now investigate how the introduced memory effects and colored noise affects the CR mechanism. For that purpose, in Fig.~\ref{CVsub_sx} we plot the CV as a function of the noise intensity. The dotted red line in plot (a) shows the result for the Markovian limit ($\Gamma=\Gamma_{\xi} \gg 1$) of Eq.~(\ref{sistema}). An interesting result is that for sufficiently long memory ($\Gamma=\Gamma_{\xi}=0.1$ case), CR is suppressed: the minimum in the $c_v \times \sigma_{\xi}$ curve is replaced by a maximum. Therefore, instead of the existence of a critical value for $\sigma_{\xi}$ such that the ISI regularity is maximized, for long memory we have a critical value of the noise intensity such that the ISI regularity is minimized. Therefore, memory is able to not only  suppress CR, but it can also induce an opposite situation, thus characterizing an ``incoherence resonance'' phenomena. 
This behavior is a manifestation of the deterministic threshold crossing induced by memory oscillations, as aforementioned. For the curves with $\Gamma=\Gamma_{\xi}=0.1\; \rm{and}\; 0.05$, the oscillation amplitude is sufficiently large such that a transition to the tonic spike regime occurs, then completely changing the qualitative aspect of the $c_v \times  \sigma_{\xi}$ curve.
It is worth noting that the authors of Ref.~\cite{lindner2002maximizing} also describe an interesting coherence-incoherence maximization mechanism regulated by the noise level in the context of a SLIF that includes an absolute refractory period.

In Fig.~\ref{CVsub_sx}(b), we call attention to the fact that, again, memory shows up as the fundamental ingredient that leads to the qualitative change in the  $c_v \times  \sigma_{\xi}$, since the characteristic time scale of the colored noise  is not able to change the qualitative aspect of the curves, nor in the $\Gamma_{\xi} \ll 1$ cases neither in $\Gamma_{\xi} \gg 1$ cases.

\section{Conclusions and final remarks}
\label{outline}

We have proposed a generalization of the stochastic resonate-and-fire model to the context of the generalized Brownian motion framework. We analyzed in detail the  consequences of the concomitant presence of memory (distributed delay) and colored noise  on the membrane potential dynamics. The model shows a remarkably rich dynamics, exhibiting a non-trivial behavior of the coefficient of variation as a function of the memory characteristic time-scale $1/\Gamma$, thus suggesting that memory can be seen as an effective mechanism for generating and controlling neuron variability. Also, the present model is able to produce ISI distributions that exhibit multimodality, a feature whose temporal pattern can be controlled by adjusting the memory time-scale, then allowing smooth transitions between unimodal and multimodal distributions. This is a very desirable mechanism, since real neurons are able to exhibit such  non-trivial patterns in the ISI distribution. Additionally, we have discussed the role played by colored noise in the membrane potential dynamics. We have shown that i) by adjusting the noise intensity it is possible to shift the  $c_v \times (\Gamma = \Gamma_{\xi})$ curve and to smooth its peak structure, and ii) by adjusting the noise time-scale $1/\Gamma_{\xi}$ a double-peak structure can emerge in the   $c_v \times \Gamma $ curve. Finally, we have studied how memory and colored noise effects modifies the establishment of the coherence resonance phenomenon. We found that long memory is able to suppress the presence of CR, and for a given range of $\Gamma$ and $\Gamma_{\xi}$, a maximum in the  $c_v \times \sigma_{\xi}$ is present instead of the usual minimization of CV that characterizes the coherence resonance. Given the variety of dynamical regimes provided by the proposed theoretical framework, we think that the present formulation could be useful to shed some new light on the modeling of dynamical aspects of real neurons. Also, the results presented here could be promptly extended to the context of other kind of noises, like the $1/f$, for example.

\begin{center}
\acknowledgements{The authors thank Fapesp for financial support (grant number 2012/24489-6).}
\end{center}


%

\end{document}